\documentstyle[psfig]{EuroPhys}
\input EuroMacr
\begin{document}
\euro{}{}{}{}
\Date{}
\shorttitle{F.S. Nogueira Gauge dependence of...}
\title{\bf Gauge dependence of the order parameter anomalous dimension 
in the Ginzburg-Landau model and the critical fluctuations in 
superconductors}

\author{F. S. Nogueira}
\institute{Centre de Physique Th\'eorique, Ecole Polytechnique, F-91128 
Palaiseau Cedex, FRANCE}

\rec{}{}
\maketitle

\begin{abstract}
The critical fluctuations of superconductors are discussed in a 
fixed dimension scaling 
suited to describe the type II regime. The gauge dependence of the anomalous 
dimension of the scalar field is stablished exactly from the 
Ward-Takahashi identities. Its fixed point value gives the $\eta$ 
critical exponent and it is shown that $\eta$ is gauge independent, as 
expected on physical grounds. In the scaling considered, $\eta$ is found 
to be zero at 1-loop order, while $\nu\approx 0.63$. This result is 
just the 1-loop values for the $XY$ model obtained in the fixed dimension 
renormalization group approach. It is shown that this $XY$ behavior holds at 
all orders. 
The result $\eta=\eta_{XY}$ should be contrasted with the negative values 
frequently reported in the literature.
\end{abstract}

The high temperature superconductors have a larger critical region relative 
to ordinary superconductors. This fact allows in principle an  
experimental access to the critical region and measurements of critical 
exponents had been made in these materials, specially the YBCO ones 
\cite{Salamon,Kamal}. 
The experiments showed that critical fluctuations are very 
important in high temperature superconductors, a consequence of this fact 
being the non-mean-field values of the critical exponents. In the tested  
temperature region it was obtained that $-0.03<\alpha<0$ and 
$\nu\approx 0.67$ while the amplitude ratio $A^{+}/A^{-}=1.065\pm 0.01$.  
These values are consistent with those found for $^4$He and correspond to a  
{\it uncharged} $d=3$ $XY$ universality class. There are some subleties 
with respect to this conclusion and the interpretation of the experimental 
data. For instance, the value of the $\nu$ exponent has been obtained by 
a direct measurement of the London penetration depth, $\lambda$, which was 
found to scale as $\lambda\sim t^{-y}$, $t$ being the reduced temperature and 
$y=0.33\pm 0.01$ \cite{Kamal}. The value of $\nu$ follows by assuming that the 
superfluid density scales with $\lambda$ as $\rho_{s}\sim\lambda^{-2}$ and 
through the use of Josephson's relation $\rho_{s}\sim\xi^{-1}$ 
\cite{Josephson}. It has been argued by Herbut and Tesanovic \cite{Herbut} 
that in a superconducting transition $\lambda$ should diverge with the 
same exponent as $\xi$, contradicting the relation 
$\rho_{s}\sim\lambda^{-2}$. In fact, they claim that the correct relation 
should $\rho_{s}\sim\lambda^{-1}$, when the critical fluctuations near the 
{\it charged} fixed point are taken into account.  
Thus, it seems to be very difficult to have an experimental access to the 
charged fixed point, even in the case of cuprates superconductors. The 
critical region probed in Refs. \cite{Salamon,Kamal} 
corresponds to a crossover regime where the gauge field 
fluctuations are unimportant. Theoretically, the nature of the charged 
fixed point has been elucidated in the early eighties through the study of 
lattice abelian gauge models using duality arguments \cite{Dasgupta}. 
It has been shown that the normal-superconducting transition should be a 
second order phase transition, at least for type II superconductors. This 
means that it must exists an infrared stable charged fixed point, 
contradicting the weak first order transition scenario of the 
$\epsilon$-expansion \cite{HLM}.     
More recent studies performed directly in continuous models 
\cite{KiometzisI,Herbut,LawrieI,deCalan} and recent numerical 
simulations in the lattice \cite{Olsson}, gives further support to this 
view.   

However, some important theoretical questions concerning the scaling 
in the Ginzburg-Landau (GL) model are still open. For instance, everybody 
agrees that the critical behavior should be the $XY$ one, but nobody is able 
to find the $XY$ value for the $\eta$ exponent from renormalization group 
calculations. More importantly, the value generally found in these 
calculations is negative. Negative values of $\eta$ does not 
violate the scaling relations provided $\eta>-1$ and, in fact, slightly 
negative values had been reported and this bound is 
fulfilled \cite{HLM,LawrieII,Radz,Folk,Herbut,Berg,deCalan}. However, the 
best $XY$ estimate is slightly positive, $\eta\approx 0.04$. Therefore, the 
negative values found are possibly an artifact of perturbation theory. It is 
worth to mention, however, that respectable values for $\nu$ 
consistent with the $XY$ behavior are 
found in some RG calculations \cite{KiometzisI,Herbut,deCalan}.  

Nearly five years ago, Kiometzis and Schakel 
\cite{KiometzisII} argued that negatives values of 
$\eta$, though $>-1$, should be unphysical since in principle it violates 
unitarity in the corresponding quantum field theory. Moreover, $\eta$ is 
the fixed point value of the anomalous dimension $\eta_{\phi}$  
of the order parameter field in the GL model, a 
quantity which is gauge dependent. Based on physical intuition only we may 
suspect that $\eta_{\phi}$ should be   
gauge independent at the fixed point. However, 
there is no actual proof of this fact to date. For this reason we shall 
address this point in this paper by using a fixed dimension RG approach for 
$T>T_{c}$. As will be made apparent soon, the critical point fixed dimension 
approach employed in references \cite{Herbut,deCalan} is not well suited 
for the following analysis. By making use of the Ward-Takahashi 
(WT) identities,  
we shall establish the gauge dependence of $\eta_{\phi}$ at all orders in 
perturbation theory. This step follows from 
general field theoretical arguments \cite{ZJ}. Next we shall show that at 
the critical point $\eta_{\phi}$ is in fact gauge independent. Finally, we 
perform a 1-loop calculation to obtain $\eta=0$, a result consistent with 
the 1-loop result for the $XY$ model. We show that higher order corrections 
will improve further this result to obtain $\eta=\eta_{XY}$. 

Our starting point is the bare action for the GL model, or 
Euclidean scalar QED in three dimensions, 

\begin{equation}
S=\int d^3x\left[\frac{1}{4}F_{0}^2+(D_{\mu}^0\phi_{0})^{\dag}
(D_{\mu}^0\phi_{0})+\frac{M_{0}^2}{2}A_{\mu}^0 A_{\mu}^0+m_{0}^2|\phi_{0}|^2
+\frac{u_{0}}{2}|\phi_{0}|^4\right]+S_{gf},
\end{equation}
where the zeroes denote bare quantities, $F_{0}^{2}$ is a short for 
$F_{0}^{\mu\nu}F_{0}^{\mu\nu}$ and $D_{\mu}^0=\partial_{\mu}+ie_{0}A_{\mu}^0$.
The $S_{gf}$ is the gauge fixing part and is given by

\begin{equation}
S_{gf}=\int d^3x\frac{1}{2\alpha_{0}}(\partial_{\mu}A_{\mu}^0)^2.
\end{equation}
We introduced a mass to the vector field in order to regularize the 
infrared divergences arising from some diagrams containing gauge fields 
propagators. This mass term breaks gauge invariance but as we shall see,  
gauge invariance is restored at the infrared stable fixed point. 

The renormalized action is defined by $S'+\delta S$ where $S'$ is the same as 
the bare action but with renormalized quantities while $\delta S$ is the 
counterterm action. It is given by

\begin{eqnarray}
\delta S&=&\int d^3x\left[\frac{Z_{A}-1}{4}F^2+(Z_{\phi}-1)(D_{\mu}\phi)^{\dag}
(D_{\mu}\phi)+(Z_{A}M_{0}^2-M^2)A_{\mu}A_{\mu}\right.\nonumber\\
&+&\left.(Z_{\phi}m_{0}^2-m^2)|\phi|^2+(Z_{u}-1)\frac{u}{2}|\phi|^4
+\frac{Z_{\alpha}-1}{2\alpha}(\partial_{\mu}A_{\mu})^2\right],
\end{eqnarray}
with the renormalized fields defined by $A_{\mu}=Z_{A}^{-1/2}A_{\mu}^0$ 
and $\phi=Z_{\phi}^{-1/2}\phi_{0}$. 

By adding sources terms for the corresponding fields, it is straightforward to 
derive the following WT identity:

\begin{equation}
\label{WT1}
\left\{\left(M^2-\frac{1}{\alpha}\Delta\right)\partial_{\mu}\frac{\delta}{
\delta J_{\mu}(x)}+ie\left[J^{\dag}(x)\frac{\delta}{\delta J^{\dag}(x)}-
J(x)\frac{\delta}{\delta J(x)}\right]\right\}W(J_{\mu},J^{\dag},J)=
\partial_{\mu}J_{\mu}(x),
\end{equation}
where $W=\log Z$, $Z$ being the generating functional of  
correlation functions. 
When the sources are zero $Z$ corresponds to the partition function. The $W$ 
generates the connected correlation functions. The Legendre transform of $W$ 
is performed as usual and 
gives the functional $\Gamma(\varphi^{\dag},\varphi,a_{\mu})$ which 
is the generator of the 1-particle irreducible functions. It satisfies a 
WT identity which is the Legendre transform of (\ref{WT1}):

\begin{equation}
\label{WT2}
\left(\frac{1}{\alpha}\Delta-M^2\right)
\partial_{\mu}a_{\mu}(x)+\partial_{\mu}\frac{\delta\Gamma}{\delta a_{\mu}(x)}+ie
\left[\varphi(x)\frac{\delta\Gamma}{\delta\varphi(x)}-\varphi^{\dag}(x)
\frac{\delta\Gamma}{\delta\varphi^{\dag}(x)}\right]=0.
\end{equation}
The WT identity given by Eq.(\ref{WT2}) gives important informations about 
the counterterms. For example, it implies that non-gauge invariant terms 
in the renormalized action are not renormalized and consenquently the 
corresponding counterterms are zero. This implies $M^2=Z_{A}M^2_{0}$ and 
$\alpha=Z_{A}^{-1}\alpha_{0}$. Gauge invariance also implies 
$e^2=Z_{A}e^2_{0}$. Let us define the following dimensionless gauge 
couplings, $\hat{e}^2=e^2/m$ and $v=m/M$. We have the following exact 
flow equations:

\begin{eqnarray}
\label{M}
m\frac{\partial M^2}{\partial m}&=&\eta_{A}M^2,\\
\label{a}
m\frac{\partial\alpha}{\partial m}&=&-\eta_{A}\alpha,\\
\label{e}
m\frac{\partial\hat{e}^2}{\partial m}&=&(\eta_{A}-1)\hat{e}^2,\\
\label{v}
m\frac{\partial v}{\partial m}&=&\left(1-\frac{\eta_{A}}{2}\right)v,
\end{eqnarray}
where we have introduced the RG function $\eta_{A}$ which is the anomalous 
dimension of the gauge field. It is defined by

\begin{equation}
\eta_{A}=m\frac{\partial}{\partial m}\log Z_{A}.
\end{equation}

An immediate consequence of (\ref{e}) is that at a charged fixed point we 
must have $\eta_{A}=1$. This very simple observation have important  
implications concerning the scaling of the magnetic field penetration 
depth, as first observed by Herbut and Tesanovic \cite{Herbut} (see also 
ref. \cite{HerbutII}). It follows also from the above equations that at the 
charged fixed points the corresponding fixed point value of $\alpha$ is 
$\alpha^{*}=0$, that is, the Landau gauge. Note that the situation here is 
somewhat different from that one encountered in particle physics where 
$d=4$. In fact, in that case the beta function for $\hat{e}^2$ is 
$\eta_{A}\hat{e}^2$ and, therefore, the charged fixed point would 
correspond to $\eta_{A}=0$. Since Eq.(\ref{a}) remains the same for $d=4$, 
we obtain that the fixed point value of $\alpha$ is arbitrary in this case. 

From Eq.(\ref{M}) we obtain that 
the charged fixed point value of $M^2$ is zero. Therefore, near the 
superconducting fixed point the effective action flows to a configuration 
with massless gauge fields in the Landau gauge. This explains why RG 
calculations performed in the Landau gauge gives good results. We can say, 
therefore, that it is legitimate to compute critical exponents in the 
Landau gauge even knowing that some RG functions like 
$\eta_{\phi}=m\partial\log Z_{\phi}/\partial m$ are gauge dependent. This 
is in contrast to $\eta_{A}$ which is gauge independent if a minimal 
subtraction scheme is used. 

In order to check the consistency of this argument 
it remains to show that $\eta_{\phi}$ is in fact well behaved 
with respect to the gauge dependence as approaching the critical point. 
We can obtain the exact gauge dependence of $\eta_{\phi}$ from the WT 
identity Eq.(\ref{WT1}). Indeed, we can use (\ref{WT1}) to relate scalar  
2-point correlation functions with a $(\partial_{\mu}A_{\mu})^2$ insertion 
to the 2-point scalar correlation functions. This is obtained by applying 
twice the WT identity (\ref{WT1}) to the 2-point correlation functions. 
We obtain in this way the exact equation 

\begin{equation}
\label{WT3}
W^{(2)}_{(\partial_{\mu}A_{\mu})^2}(p)=2e^2\int\frac{d^3k}{(2\pi)^3}
\frac{\alpha^2}{(k^2+\alpha M^2)^2}\left[W^{(2)}(p+k)-W^{(2)}(p)\right], 
\end{equation}
out of which we get the {\it exact} gauge dependence of $\eta_{\phi}$:

\begin{equation}
\label{gaugedep}
\frac{\partial\eta_{\phi}}{\partial\alpha}=-\frac{\eta_{A}\hat{e}^2 v}
{8\pi\sqrt{\alpha}}, 
\end{equation}
with $\eta_{A}$ being gauge independent. As a charged fixed point is 
approached, $\eta_{A}\to 1$, $v\to 0$, $\hat{e}\to\hat{e}^{*}$ 
and $\alpha\to 0$. In order to have a consistent $\eta$ exponent it is 
necessary to show that $v/\sqrt{\alpha}\to 0$ as we approach the 
critical point. This is in fact the case since from Eqs.(\ref{a}) and 
(\ref{v}) we obtain that $v/\sqrt{\alpha}$ scales like $m=\xi^{-1}$ near 
the critical point. Therefore, we have 
$\partial\eta_{\phi}/\partial\alpha=0$ at the critical point. The same 
result is obtained if we consider a more general model including a 
Chern-Simons term whose critical behavior has been studied recently
\cite{deCalan,Kleinert,Malbouisson}. 
Although in this case the photon has a massive 
propagator without breaking of gauge invariance, we must still keep the 
mass $M$ above in order to recover consistently the non-topological model 
in the zero Chern-Simons mass limit. Note that the critical point approach 
\cite{Herbut,deCalan}, though it regularizes infrared divergent graphs, is 
not appropriated to discuss in a complete way the gauge dependence of 
$\eta_{\phi}$. In fact, at the critical point $m=M=0$ and we have infrared 
divergences in (\ref{WT3}). 

Let us perform now a sample calculation up to 1-loop order in the Landau 
gauge. For this end we write $\phi=(\phi_{1}+i\phi_{2})/\sqrt{2}$ and use the 
the renormalization conditions for the irreducible vertex functions: 
$\Gamma^{(2)}_{11}(0)=m^2$, $\Gamma^{(2)}_{\mu\mu}=3M^2$, 
$\partial\Gamma^{(2)}_{11}(0)/\partial p^2=1$, 
$\partial\Gamma^{(2)}(0)_{\mu\mu}/\partial p^2=2$ and 
$\Gamma^{(4)}_{1111}(0)=3u$. 
The corresponding anomalous dimensions are given by 

\begin{eqnarray}
\eta_{A}&=&\frac{\hat{e}^2}{24\pi},\\
\eta_{\phi}&=&-\frac{2}{3\pi}\frac{\hat{e}^2v^2}{(1+v)^2}.
\end{eqnarray}
The flow of the coupling $\hat{u}=u/m$ is given up to 1-loop order by 

\begin{equation}
m\frac{\partial\hat{u}}{\partial m}=(2\eta_{\phi}-1)\hat{u}+\frac{5}{8\pi}
\hat{u}^2+\frac{v}{2\pi}\hat{e}^4.
\end{equation}
Figures 1 and 2 show the flow diagram respectively in the $(u,f)$ 
($f\equiv\hat{e}^2$) and the $(f,v)$ planes. Fig.1 corresponds to a section 
$v=0.001$ of the critical manifold. Note that for a small but nonzero $v$ 
we have two charged fixed points, corresponding respectively to the 
tricritical and superconducting fixed points \cite{Herbut,Berg,deCalan}. 
It is useful to compare the above calculation with other fixed dimension 
approaches, for instance, the one employed by Herbut and Tesanovic and 
de Calan {\it et al.} \cite{Herbut,deCalan}. In the approach of Refs. 
\cite{Herbut,deCalan} the charged fixed points are obtained in a critical 
point calculation through the introduction of a constant parameter $c$, 
giving the ratio between two different momentum scales of the problem, 
namely, the momentum determining the running of $\hat{e}^2$ and the one 
determining the running of $\hat{u}$. The parameter $c$ can be adjusted 
in order to generate charged fixed points. The arbitrariness of $c$ is 
removed by fixing it from a known numerical value of the Ginzburg constant, 
$\kappa$, at the tricritical fixed point. In our case, it is $v$ that plays 
the role of $c$, $v$ representing the ratio between the two existing scales 
in our problem, namely, $m$ and $M$. An important difference between the 
present approach and the one of Refs. \cite{Herbut,deCalan} is that we do 
not need to fix numerically $v$ since it flows naturally to a fixed point 
value. Note that the infrared stable fixed point in the flow diagram of 
Fig.2 is charged. 
The problem with the $\epsilon$-expansion is that only one scale is 
considered. In such a model we have naturally two scales, which are of course 
related. For instance, the most natural scales in the broken symmetry 
phase are $\lambda$ and $\xi$, whose ratio gives an important physical 
parameter, the Ginzburg constant $\kappa$. However, the $\epsilon$-expansion 
has the advantage of being a controlled approximation, in the sense that 
we have a well defined small parameter. In the fixed dimension approach, a 
good choice of expansion parameter is $1/N$, $N$ being the number of order 
parameter components. The main drawback in this case is that it is not 
easy to extrapolate the value of $N$ to the physical case $N=2$. Thus, 
our approximation, though uncontrolled (just like those in Refs. 
\cite{Herbut,deCalan,Berg}, enables us to get sensible physical results. 
Anyway, it is possible in principle to use controlled approximations like  
the $\epsilon$-expansion to obtain results consistent with the existence of 
an infrared stable charged fixed point. This can be accomplished by 
considering explicitly the two scales of the model. This problem is 
treated more appropriately in the broken symmetry regime and will be the 
subject of a future publication.      

\begin{figure}
\centerline{\psfig{figure=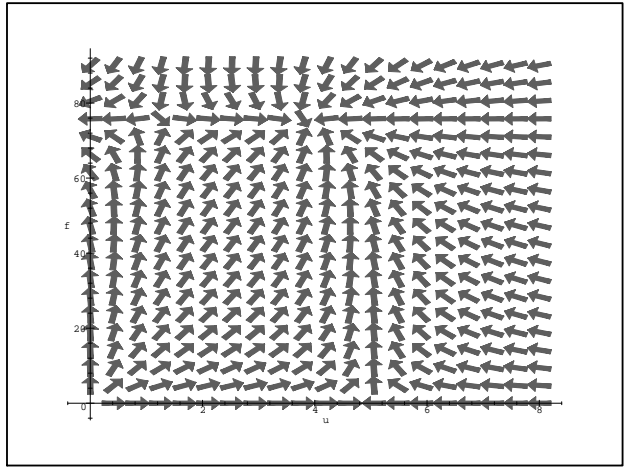,height=7truecm,angle=-90}}
\caption{Flow diagram in the $(u,f)$-plane.}
\end{figure}

\begin{figure}
\centerline{\psfig{figure=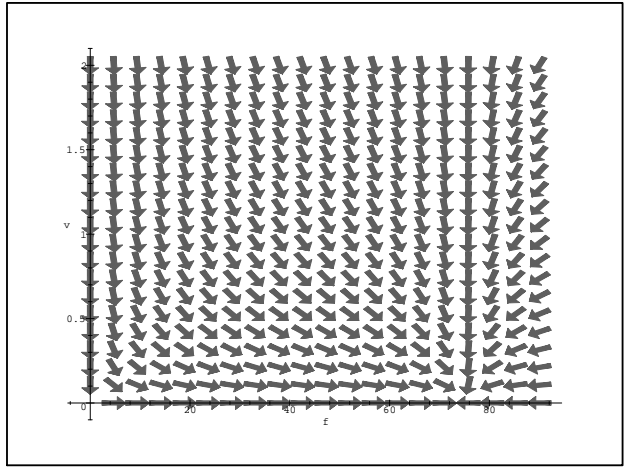,height=7truecm,angle=-90}}
\caption{Flow diagram in the $(f,v)$-plane. The infrared stable fixed 
point is charged}
\end{figure}

Since $v=0$ at the charged fixed point, we obtain exactly the 
critical behavior of the $XY$ model. For instance, we find 
$\eta=0$ and $\nu\approx 0.63$, the 1-loop values for $XY$ model in the 
fixed dimension approach. Note that $\eta$ is not negative in the present 
scaling. Higher order corrections behave in the same way, that is, all the 
powers of $\hat{e}^2$ are suppressed at the charged fixed point because 
they are multiplied by some function of $\hat{v}$ which goes to zero at the 
charged fixed point. Note that these functions of $\hat{v}$ will never be 
divergent as $\hat{v}\to 0$. This follows by simple dimensional analysis 
performed in the graphs containing gauge and scalar 
fields lines. Thus, if we compute 
higher order corrections we expect to approach asymptotically the best 
$XY$ values estimates for $\eta$ and $\nu$.  

In summary, we stablished exactly the gauge dependence of the scalar field 
anomalous dimension and showed that calculations are legitimate if 
performed in the Landau gauge. In the fixed dimension scaling considered, 
explicit calculations show that the GL model lies in fact in the $XY$ model 
universality class. 

\newpage

\stars

The author would like to thank C. de Calan for interesting discussions. The 
idea of writing this paper originated from discussions with 
R. Folk, Yu. Holovatch, D. Loison and A.M.J. Schakel and the author is 
particularly indebted to them. This work was supported by the agency CNPq, a 
division of the Brazilian Ministry of Science and Technology.

\end{document}